# Influence of Device Geometry on Transport Properties of Topological Insulator Microflakes*

Fan Gao(高凡)[1,2], Yongqing Li(李永庆)[1,2,3,4]**

[1]*Beijing National Laboratory for Condensed Matter Physics, Institute of Physics, Chinese Academy of Sciences, Beijing 100190, China*
[2]*School of Physical Sciences, University of Chinese Academy of Sciences, Beijing 100190, China*
[3]*Songshan Lake Materials Laboratory, Dongguan, Guangdong 523808, China*
[4]*CAS Center for Excellence in Topological Quantum Computation, University of Chinese Academy of Sciences, Beijing 100190, China*

*Supported by the National Natural Science Foundation of China (Project No. 61425015), the Strategic Priority Research Program of Chinese Academy of Sciences (Project No. XDB28000000), and the National Key Research and Development Program (Project No. 2016YFA0300600).
**Corresponding author. Email: yqli@iphy.ac.cn

Telephone: (+86)13911341236
E-mail: f.gao@iphy.ac.cn; yqli@iphy.ac.cn

**Abstract**

In the transport studies of topological insulators, microflakes exfoliated from bulk single crystals are often used because of the convenience in sample preparation and the accessibility to high carrier mobilities. Here, based on finite element analysis, we show that for the non-Hall-bar shaped topological insulator samples, the measured four-point resistances can be substantially modified by the sample geometry, bulk and surface resistivities, and magnetic field. Geometry correction factors must be introduced for accurately converting the four-point resistances to the longitudinal resistivity and Hall resistivity. The magnetic field dependence of inhomogeneous current density distribution can lead to pronounced positive magnetoresistance and nonlinear Hall effect that would not exist in the samples of ideal Hall bar geometry.

PACS: 73.63.-b, 73.20.-r, 73.43.Qt

Topological insulators (TIs) have emerged as an important class of materials for pursuing exotic quantum phenomena as well as novel device applications.[1, 2] However, it has been very challenging to obtain high-quality 3D TI samples for electron transport experiments, despite a lot of efforts on the growth of high-quality TI materials.[3, 4] Owing to remarkable advances in the device fabrication techniques for Van der Waals materials,[5] TI microflakes exfoliated from bulk single crystals (e.g. $(Bi,Sb)_2(Te,Se)_3$ and Sn-doped $(Bi,Sb)_2(Te,S)_3$) have shown some advantages over TI thin films and nanoplates grown with molecular beam epitaxy or chemical vapor deposition.[6-13] By using exfoliated samples, well-defined quantum Hall plateaus have been observed in the magnetic fields lower than 5 T.[8, 9] The flake-shaped samples allow for straightforward assembly of dual-gated devices by utilizing hexagonal boron nitride flakes as dielectric layers.[7, 10-13] In order to achieve the lowest possible disorder in the surface states, the TI flakes are often not processed into standard Hall bars. Even though such practice is very common in the studies of 2D materials, the influence of metal electrodes, which typically overlap with conducting path of the microflakes by a few microns, may not be negligible for evaluation of the transport parameters. Moreover, the spatially inhomogeneous current flow caused by the electrode configuration is further complicated by coexistence of the surface and bulk states in 3D TIs and the applied magnetic field. To the best of our knowledge, no systematic work has been reported to quantitatively investigate the effects of device geometry on the transport measurements.

In this letter, we report numerical simulation results on the impact of electrical contacts on 3D TI samples with non-Hall bar structures. We show that a geometry correction factor needs to be introduced to accurately determine the longitudinal resistivity from the measured four-point resistance in zero magnetic field. When a perpendicular magnetic field is applied, the inhomogeneous current distribution is modified, resulting in positive magnetoresistances and nonlinear Hall effects that would be absent for ideal Hall-bar-shaped samples.

Our numerical simulation is based on the semiclassical transport theory in the diffusive regime, in which the current density and electrical field satisfy the Ohm's law $J = \sigma E$. When the magnetic field is applied along the z axis, the conductivity tensor $\sigma$ can be written as

$$\sigma = \frac{\sigma_0}{1+(\mu_H B)^2} \begin{pmatrix} 1 & R_H B \sigma_0 & 0 \\ -R_H B \sigma_0 & 1 & 0 \\ 0 & 0 & 1+(\mu_H B)^2 \end{pmatrix}. \quad (1)$$

Here $\sigma_0$ is the zero-field longitudinal conductivity, $\mu_H$ is the Hall mobility, and $R_H$ is the Hall coefficient (see the Supplemental Information for details). The electrical field distribution inside the sample can be obtained by solving the Laplace equation

$$\nabla^2 \phi = 0, \quad (2)$$

and subsequently applying $E = -\nabla \phi$. In solving Eq. (2), boundary conditions $J \cdot \hat{n} = 0$ and $E \cdot \hat{t} = 0$ are applied for insulating and conducting boundaries, respectively, where $\hat{n}$ is the unit normal vector and $\hat{t}$ is the unit tangential vector.[14]

In the TI samples based on exfoliated microflakes, the metallic electrodes are usually in contact with either the top or the bottom surface. Both surfaces, as well as the bulk states, can participate in electron transport. In our finite element analysis, the surface states are simulated by a 1 nm thick layer with a conductivity much higher than the bulk, and the electrical contacts are only in touch with the bottom surface. By varying the bulk and surface conductivities, electrode configuration, and magnetic field, the influence of device geometry on the transport can be investigated systematically.

Sketched in Fig. 1(a) is a four-terminal device configuration often used in the measurements of exfoliated microflakes.[15-21] Here, the TI microflake with insulating bulk is modeled as a 20 μm wide and 100 nm thick rectangular slab, with bulk and surface conductivities of 0 and 0.25 mS, respectively. The metal electrodes are 5 μm wide, separated by 20 μm to each other, and have a sheet conductivity of 250 mS. A current of 1 μA is applied to the sample via the source and drain electrodes (I$^+$ and I$^-$). Fig. 1(b) shows the calculated electric potential distribution on the top surface in zero magnetic field. The corresponding current densities are obtained by applying Ohm's law. Fig. 1(c) displays the variation of $J_x$, the x component of the current density, across the conducting channel (line W in Fig. 1(b)). In contrast to nearly constant current density expected for a homogeneously conducting thin film, the current density decreases as the point of interest moves from the sample edge to the center. In this particular case, the spatial inhomogeneity is associated with the insulating bulk, which forces the current to flow from the bottom to the top surface via the sample edges. Such a lateral current component is consistent with the spatial distribution of $J_x$ along the same line on the bottom surface. As depicted in Fig. 1(c), $J_x$ has a maximum in the center, with an average magnitude larger than that of the top surface.

The influence of metal electrodes can be seen more clearly in Fig. 1(d), which displays the variations of $J_x$ in the longitudinal direction (line L in Fig. 1(b)) for both surfaces. One striking feature is the oscillations in current density. The oscillating magnitude on the bottom surface is much stronger than that on the top surface. This can be attributed to the electrodes in direct contact with the bottom surface. The current shunting effect of the electrodes is responsible for the zero $J_x$ plateaus observed for the bottom surface. The other noticeable feature in Fig. 1(d) is that the difference in $J_x$ between the two surfaces also varies with the distance to the electrodes. This suggests the existence of an inter-surface current component. For the samples with insulating bulk, such a lateral current component must go through the side surfaces, complicating the extraction of the transport parameters (e.g. longitudinal and transverse resistivities) from the measured resistances.

The asymmetric current distribution discussed above makes the measured four-point resistances larger than their counterparts in the Hall bars with the same channel widths and electrode spacings. For the latter, the surface resistivity can be evaluated straightforwardly from the aspect ratio, $\rho_{\text{surf}} = R_{xx,\text{H}} / \left(\frac{l}{2(w+d)}\right)$ (see the Supplemental Information), in which the 'H' in the subscript denotes the standard Hall bar geometry and the factor '2' in the denominator takes account of equal contributions of the top and bottom surfaces as well as those of the two side surfaces. For the device of non-Hall-bar configuration, a geometry correction factor must be introduced to obtain the correct surface resistivity. It is defined as $f_\text{L}=R_{xx,\text{N}}/R_{xx,\text{H}}$, the ratio between the four-point resistances in the non-Hall bar and standard Hall bar geometries. For the bulk-insulating TI device depicted in Fig. 1(a), our numerical result is $f_L$=1.09, a small but not negligible correction for the quantum transport experiments that require quantitative values of conductivities.[22-24]

For most of the 3D TIs encountered in experiments, the transport is further complicated by the bulk conductivity. Even for $(Bi,Sb)_2(Te,Se)_3$, a TI material reported to have the most insulating bulk to date, the resistivity is only up to about 10 Ω·cm.[25-27] The finite bulk resistivity creates additional paths for the inter-surface current, but nonetheless decreases the asymmetry effect caused by the one-sided electrical contacts. As shown in Fig. 2(a), the geometry correction factor drops from $f_L$=1.09 at the bulk-insulating limit toward 1.00 as the bulk resistivity decreases. In fact, for $\rho_{bulk}$< 0.1 Ω·cm, $f_L$ is very close to 1, and is no longer sensitive to the resistivity. It should be pointed out, however, that the $f_L$ values at the two resistivity limits are not universal — they depend on the details of device parameters. It can be seen in Fig. 2(a) that $f_L$ varies weakly with the sample thickness. In contrast, the geometrical correction factor exhibits much stronger dependences on other sizes, including electrode spacing ($l$), channel width ($w$), and electrode width ($l_V$), as depicted in Fig. 2(b-d). In the case of high bulk resistivity, $f_L$ can exceed 1.2 for some device parameters. One general trend is that $f_L$ values close to 1 can be obtained in the samples with large $l/w$ ratios and narrow electrode widths. This can serve as a guiding principle for future sample design. Nevertheless, it is noteworthy that for the samples with low bulk resistivities, the geometry correction effect is rather weak, unless the $l/w$ ratio is very small. This can help simplify the data analyses in many transport measurements of exfoliated microflakes.

The geometrical corrections can be modified significantly by the magnetic field. Even in a very simple case, such as a multiterminal device based on a 2D material with a single type of charge carriers, the interplay between the magnetic field and electrical contact configuration can produce geometry-related magnetoresistances and nonlinear Hall effect. These effects would be absent for the samples with the Hall bar geometry. The bottom panel of Fig. 3(a) shows the distribution of electric potential $\phi$ for a six-terminal 2D sample with a channel width of 20 μm and the Hall angle $\varphi_H$ satisfying $\tan \varphi_H = \rho_{xy}/\rho_{xx} = 1$. The equipotential lines far away from the electrodes are nearly straight. In comparison with the nearly vertical lines in the zero-field case (Fig. 1(b)), they are tilted by about 45° due to the applied magnetic field. Near the electrodes, the equipotential lines are clearly bent. This can be understood as a proximity effect of the highly conductive electrodes, which make the electric potential nearly constant in the contact regions, as demonstrated in the top panel of Fig. 3(a) by variations of $\phi$ with the $y$-coordinate along several vertical lines. The influence of electrodes is also manifested in the vector plot of current density $\mathbf{J}(x,y)$ for the same Hall angle (Fig. 3(b)).

As the magnetic field gets stronger, the Hall angle increases, and consequently the magnetic field competes more strongly with the proximity effect of electrodes. This leads to larger geometrical corrections in the transport properties. Fig. 3(c) and (d) show that the geometrical correction factors for the longitudinal and Hall resistances, $f_L$ and $f_T$, increase with the Hall angle, where the latter correction factor is defined as $f_T = R_{yx,N}/R_{yx,H}$, similar to the definition of $f_L$. The magnitudes of $f_L$ and $f_T$ depend on specific parameters of the sample geometry. In general, the devices with larger overlaps with the electrodes tend to have greater geometrical correction factors. For instance, for the device with $l_V = 5$ μm and $w_V = 0$, $f_L$ exceeds 2.2 for $\tan \varphi_H = 2$. Here, $w_V = 0$ represents a limiting case that the device degrades to the four-terminal geometry (Fig. 1(a)). Increasing $w_V$ to 10 μm reduces the contact area of the voltage electrodes by one half, and $f_L$ decreases to 1.34 for the same Hall angle. Further decrease in the contact area of the electrodes by increasing $w_V$ and reducing $l_V$ can make the geometrical correction negligible (e.g. $f_L$ remains close to 1 for $l_V = 0.1$ μm and $w_V = 18$ μm, see Fig. 3(c)). The correction to the Hall resistance follows the similar trend. As shown in Fig. 3(d), the $f_T$ values for all three sets of device parameters are smaller than 1, and the device with the smallest electrode width and the largest electrode splitting ($l_V = 0.1$ μm and $w_V = 18$ μm) has the smallest deviation from the standard Hall bar.

According to the Drude model, the longitudinal resistance is independent of the magnetic field in a 2D system with single carrier type. The magnetic field dependence of $f_L$ (Fig. 3(c)), however, means the existence of positive magnetoresistances, as shown in Fig. 3(e). Similarly, the field dependence of $f_T$ causes the deviation of the Hall resistance from the conventional linear response. As displayed in Fig. 3(f), the nonlinear Hall effect is most pronounced for the device with the largest electrical contacts ($l_V$=5 μm and $w_V$=10 μm).

The magnetoresistances and nonlinear Hall resistances discussed above originate from the interplay between the magnetic field and device geometry. These effects are therefore not restricted to the 2D systems alone. In Fig. 4 we show the numerical results for a six-terminal 3D TI device with an insulating bulk and $l_V$=5 μm, $w_V$=10 μm, $w$=20 μm. As depicted in Fig. 4(a) and (b), the $f_L$ and $f_T$ values deviate strongly from 1. Plotted in Fig. 4(c) and (d) are the magnetoresistances and Hall resistances of this device obtained with $\rho_{surf}$ = 4 kΩ and $R_H$ = 1 kΩ/T for both the top and bottom surfaces. These transport parameters correspond to a carrier density of about 6.2×10$^{11}$ cm$^{-2}$, and a mobility of 2.5×10$^3$ cm$^2$V$^{-1}$S$^{-1}$, both within experimental reach in 3D TIs. It is also worth noting that observations of positive magnetoresistances and nonlinear Hall effects were not rare in previous studies of 3D TI samples.[23, 28] They were often attributed to coexistence of multiple types of carriers on the two surfaces and in the bulk. The results of this work suggest that a careful differentiation from the geometrical correction effects need to be made in order to acquire reliable results on a quantitative level, such as the extraction of the carrier densities and mobilities based on the two-band model,[23, 29, 30] and the precise evaluation of longitudinal magnetoconductivity for the study of quantum corrections from weak antilocalization and electron-electron interaction effects.[22-24]

In conclusion, we have carried out a numerical study of electric potential and current density distributions in 3D TI devices with non-Hall-bar geometries. We have demonstrated that both the longitudinal and Hall resistances could deviate strongly from those of the standard Hall bars based on the same material and with the same aspect ratios. Geometrical correction factors must be introduced to extract correct values of longitudinal and Hall resistivities. It is also found that the correction factors are sensitive to the device geometry, transport parameters of the surface and bulk states, and the magnetic field. The dependences of the geometrical correction factors on the magnetic field lead to the positive magnetoresistances and nonlinear Hall effect that would be absent for an ideal Hall bar sample based on the same TI material. Our work calls for caution in the study of transport properties of topological materials, in particular those with highly conductive surface states.

**Figures & captions:**

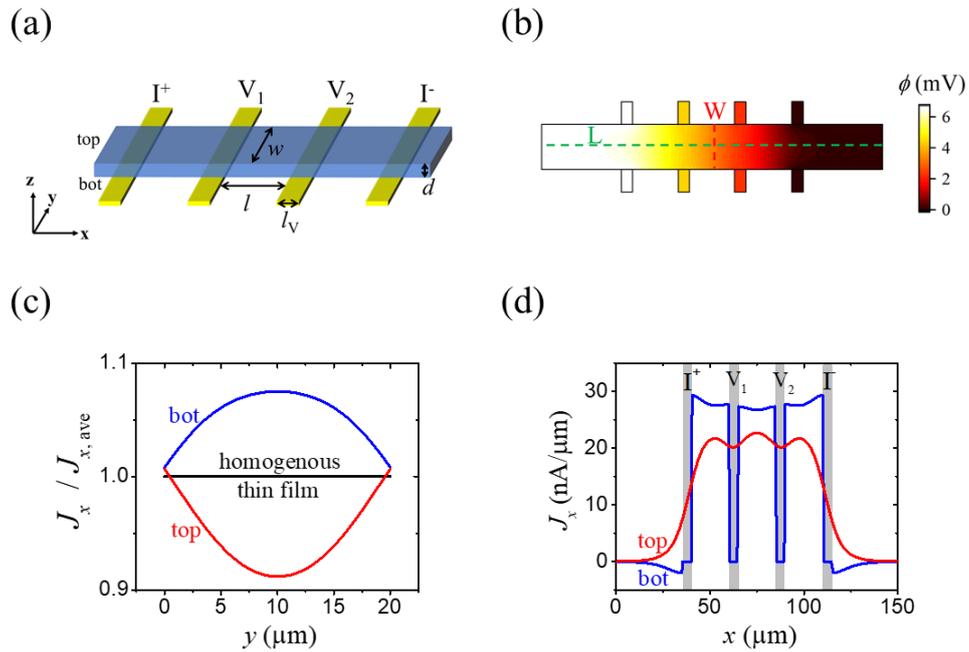

Fig. 1 Numerically calculated distributions of electric potential and current density in a TI device with insulating bulk. (a) Sketch of a 4-terminal device with electrodes located underneath the TI microflake. (b-d) Numerical results for a device with $w$=20 μm, $l$=20 μm, $l_V$=5 μm, $d$=100 nm, $I$=1 μA, and $\rho_{surf}$=4 kΩ. (b) Potential distribution on the top surface. (c) Variations of $J_x$ in the lateral direction (along line W in panel (b)) for the top surface (red) and bottom surface (blue). The black line represents the $J_x$ distribution in a homogenously conducting thin film. All $J_x$ values are normalized for clarity. (d) Variation of $J_x$ in longitudinal direction (along line L in panel (b)). The gray areas denote the locations of electrical contacts.

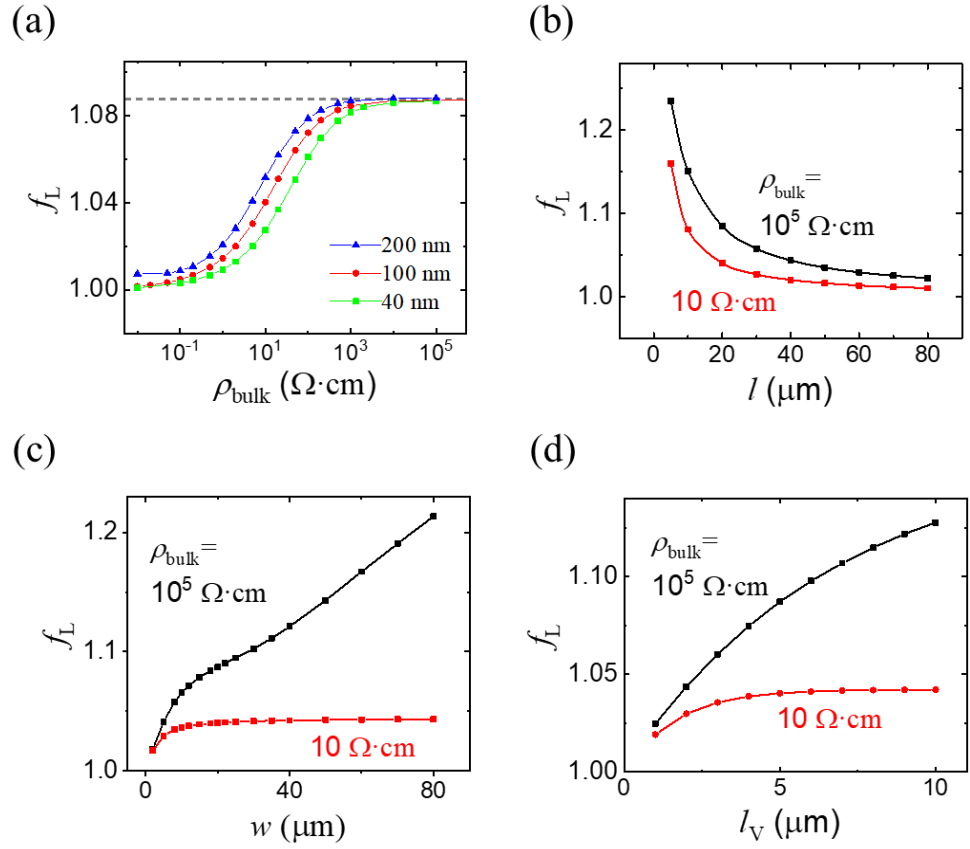

Fig. 2 Geometrical correction factors for the longitudinal resistance in a 4-terminal TI device. (a) Dependences of $f_L$ on the bulk resistivity for the TI flakes with thicknesses of 40, 100 and 200 nm. (b-d) Dependences of $f_L$ on electrode spacing (b), channel width (c), and electrode width (d). The geometry parameters involved here are defined in Fig. 1(a). All geometrical and transport parameters are same as those used in Fig. 1, except those specified in each panel.

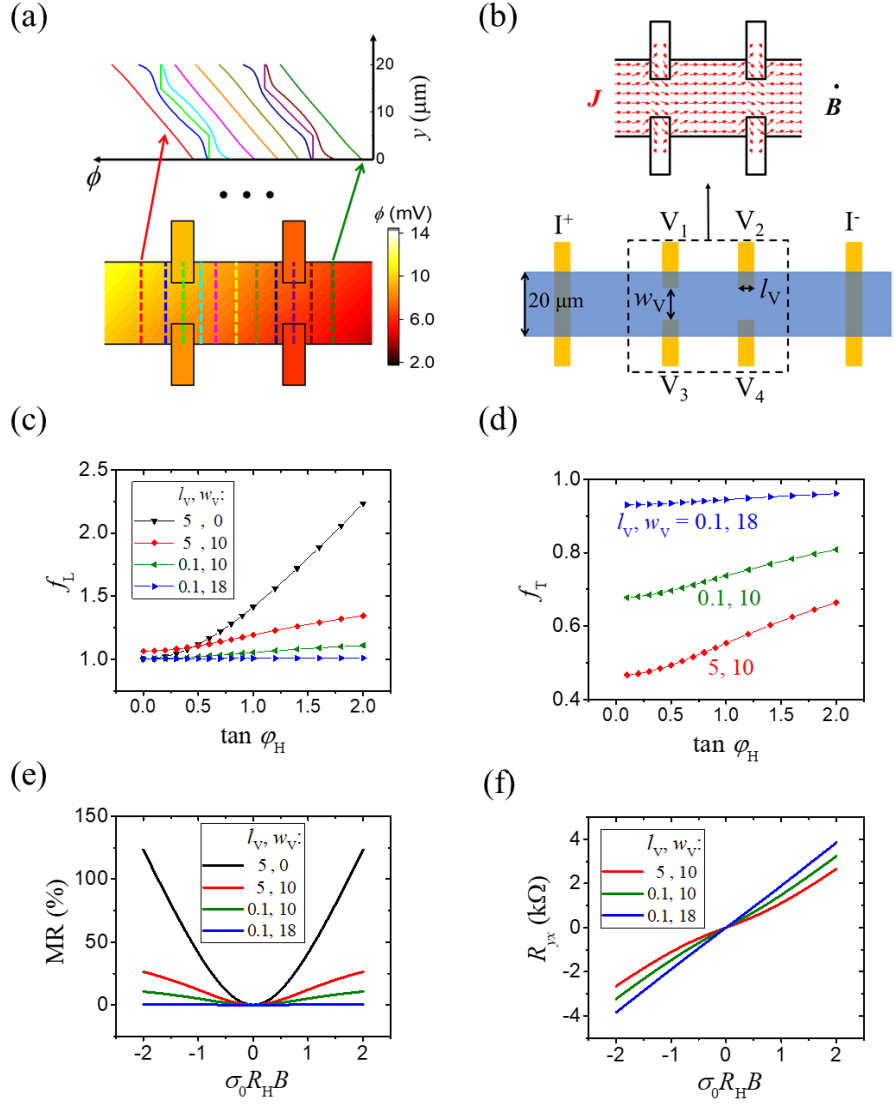

Fig. 3 Influence of magnetic field on the geometrical correction factors for a six-terminal device based a 2D system with a single type of carriers. (a) Electric potential distribution for $\tan\varphi_H=1$. Curves in the upper panel correspond to the vertical dashed lines in the lower part. (b) Vector plot of the current density (upper) and layout of the device (lower). The Hall angle satisfying $\tan\varphi_H=1$ is used for numerical simulations shown in panels (a) and (b). (c, d) Dependences of geometrical correction factors $f_L$ (c) and $f_T$ (d). (e, f) Magnetoresistances (e) and Hall resistances (f) corresponding to the geometrical corrections shown in panels (c) and (d). In panels (c)-(f), the $l_V$ and $w_V$ values are in unit of μm, and a sheet resistivity of 2 kΩ is used for the simulation.

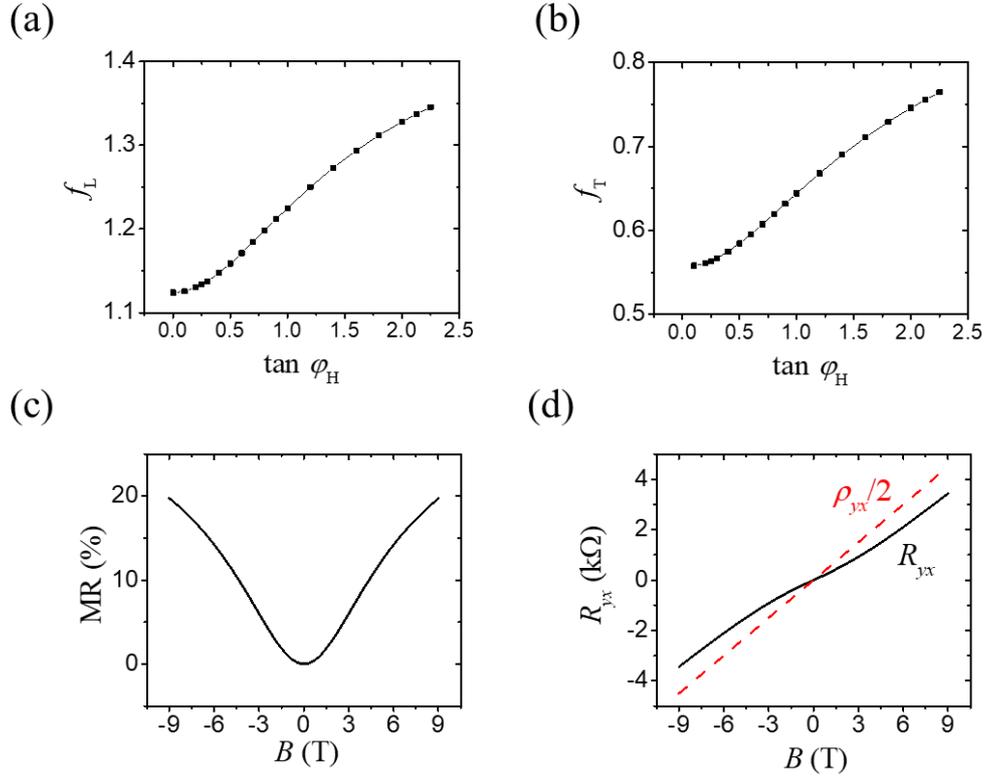

Fig. 4 Magnetoresistance and nonlinear Hall effect arising from geometrical corrections in a 3D TI sample. The device geometry is similar to the 6-terminal device shown in Fig. 3(b). A 20 μm wide rectangular TI microflake with a thickness of 100 nm is placed on the metal electrodes with $l_V$=5 μm and $w_V$=10 μm. (a, b) Dependences of $f_L$ (a) and $f_T$ (b) on tan $\varphi_H$. (c, d) Magnetoresistance and nonlinear Hall effect curves corresponding to the geometrical correction factors shown in panels (a) and (b). For each of the top and bottom surfaces, $\rho_{xx}$ = 4 kΩ, $\rho_{yx}$ = $R_H B$, and $R_H$ = 1 kΩ/T are used for the numerical calculations.